\documentstyle[preprint,prl,aps,tighten]{revtex}

\newcommand\struct\relax

\begin{document}
\draft

\title{Berry phase, hyperorbits, and the Hofstadter spectrum}
\author{Ming-Che~Chang and Qian~Niu}
\address{Department of Physics, University of Texas at Austin, Austin, TX
78712}
\date{\today}
\maketitle

\begin{abstract}
We develop a semiclassical theory for the dynamics of electrons
in a magnetic Bloch band, where the Berry phase plays an important role.
This theory, together with the Boltzmann equation, provides a framework
for studying transport problems in high magnetic fields.
We also derive an Onsager-like formula for the quantization of cyclotron
orbits,
and we find a connection between the number of orbits and Hall
conductivity. This connection is employed
to explain the clustering structure of the Hofstadter spectrum.
The advantage of this theory is its generality and conceptual simplicity.
\end{abstract}
\pacs{PACS numbers: 72.10.--d 72.15.Gd 73.20.Dx}
\narrowtext

The theory of semiclassical dynamics of Bloch
electrons in a weak electromagnetic field plays a
fundamental role in our understanding of electronic spectral and
transport properties in metals and semiconductors.
The basic ingredients of this theory are the following
pair of equations:
\begin{equation}
{\bf\dot{r}}= \frac{\partial {\cal E}_n({\bf k})}{\hbar\partial {\bf k}},\ \
\hbar{\bf\dot{k}}= -e{\bf E}-e{\bf\dot{r}}\times{\bf B},
\end{equation}
where ${\cal E}_n({\bf{k}})$ is the energy for the $n$-th band,
and ${\bf{E}}$ and ${\bf{B}}$ are the electric and
magnetic fields \cite{Mermin}.
The validity of these relations depends on the absence
of interband tunneling, therefore (1) holds if the
external fields are sufficiently weak.

A natural question is how should these semiclassical equations be modified for
a
magnetic Bloch band (MBB). Such a band is obtained when an electron is
subject simultaneously to a periodic potential and a magnetic field
(not necessarily weak),
such that the magnetic flux per unit cell  of the periodic potential
(plaquette)
is a rational multiple of the flux quantum $h/e$.  For example,
in a tight binding model Hofstadter showed that a Bloch band is
broken into $q$ subbands if the rational number is $p/q$ \cite{Hofstadter}.
At the opposite limit, when the magnetic field
is much stronger than the periodic potential, a Landau level is broadened and
split
into $p$ subbands \cite{Rauh}.
These subbands at both limits are manifestations of the magnetic Bloch bands.

It is difficult to resolve these MBBs in a naturally occuring solid.
For example, with a lattice constant $a=5\ $\AA\  and a magnetic field $B=3$
Tesla,
the value of $p/q$ is of order $10^{-2}$.
However, $p/q$ can be a significant fraction of unity
if we are using an artificial lattice with a
much larger period.
This is realized, for example, by imposing an optical interference pattern
on top of a heterostructure to create a grid potential
on the two dimensional electron gas beneath.
Recent experiments have demonstrated that disorder can be reduced
to such a degree that the effect of magnetic bands emerges
in the transport properties of the electron gas \cite{Gerhardts}.

Our goal is to derive the counterpart equations to Eq.~(1) for a MBB, and use
them to explore
the dynamics of electrons under a weak electromagnetic perturbation.
To simplify the discussion, assume the electrons are confined in a
two-dimensional periodic potential, with a magnetic field perpendicular to it.
Because the vector potential for a
constant magnetic field is not periodic, the Hamiltonian
$H$ does not commute with usual translation operators.
However, we can define {\it magnetic} translation operators $\tilde T({\bf R})$
that commute with $H$ \cite{Zak}. Analogous to the Bloch states,
we require the eigenstates of $H$ to satisfy the relation
$\tilde T({\bf R})\Psi_n({\bf k})=\exp(i{\bf k}\cdot{\bf R})\Psi_n({\bf k})$
\cite{Chang}.
This is not well-defined in general, because the magnetic translation operators
for different displacements, ${\bf R}_1$ and ${\bf R}_2$, do not commute
unless there is an integer number of flux quanta in the area $|{\bf
R}_1\times{\bf R}_2|$.
Therefore, if $\phi=p/q$, we have to choose a unit cell consisting of $q$
plaquettes.
Correspondingly, the area of the magnetic Brillouin zone is reduced by a factor
of $q$.
Furthermore,
because of the magnetic translation symmetry, the energy spectrum is exactly
$q$-fold degenerate.

\vskip .1in
\noindent{\it Electric perturbation and transport ---}
If we write a magnetic Bloch state in the form $\Psi_n({\bf k}_0)=
\exp(i{\bf k}_0\cdot{\bf r})u_n({\bf k}_0)$, then
$u_n$ is modified by a weak and homogeneous electric field into
\begin{equation}
u_n({\bf k})-
i\hbar\sum_{n'\neq n}
\frac{ u_{n'}({\bf k})\langle u_{n'}({\bf k})|
{\dot u}_{n}({\bf k})\rangle } { {\cal E}_n({\bf k})-{\cal E}_{n'}({\bf k}) },
\end{equation}
where ${\bf k}={\bf k}(t) \equiv {\bf k}_0-e{\bf E}t/\hbar$, and we have used a
time
dependent vector potential for the electric field.  The above result is
obtained
by adiabatic perturbation theory, which is valid for weak fields.
The average velocity in such a state can be easily evaluated as:
\begin{equation}
{\bf{\dot r}}=\frac{\partial {\cal E}_n({\bf k})}{\hbar\partial {\bf k}}-
{\bf{\dot k}}\times {\hat z}\ \Omega_n({\bf k}),\ \
\hbar{\bf{\dot k}}=-e{\bf E},
\end{equation}
where ${\hat z}$ is the unit vector along the direction of the magnetic field.
The second term in the expression for ${\bf\dot r}$ comes from the first
order non-adiabatic correction in the wave function, with \cite{TKNdN}
\begin{equation}
\Omega_n({\bf k})=i\left( {\langle \frac{\partial u_n}{\partial k_1}|
\frac{\partial u_n}{\partial k_2}\rangle-
\langle \frac{\partial u_n}{\partial k_2}
|\frac{\partial u_n}{\partial k_1}\rangle } \right).
\end{equation}

The equations in Eq.~(3) are the new set of semiclassical equations.
Notice that the usual Lorentz force term  for ${\bf\dot k}$ is absent because
the magnetic field has already been included in the band
structure.  On the other hand, the velocity has an extra term involving
$\Omega_n({\bf k})$, which will be called the ``curvature" of the Berry phase,
because its integral over an area bounded by a path $C$
in ${\bf k}$-space is the Berry phase $\Gamma_n(C)$ \cite{Berry}.
 In physical terms, $\Omega_n({\bf k})$ describes the contribution
of state $\Psi_n({\bf k})$ to the Hall conductivity in the
absence of scattering.
The derivation of Eq.~(3) is based on a homogeneous
field; nevertheless, it should still be valid when the field is slowly varying
in space and time.
The generalization of Eq.~(3) to higher dimensions is straightfoward. In
that case there will be more than one component of the curvature.

The combination of Eq.~(3) with the Boltzmann equation,
\begin{equation}
{\bf{\dot r}}\cdot \frac{\partial f}{\partial {\bf r}}+
{\bf{\dot k}}\cdot \frac{\partial f}{\partial {\bf k}}=
\left( \frac{\partial f}{\partial t}\right)_{\rm coll},
\end{equation}
offers a general framework for {\it semiclassical}
transport in a MBB. The right hand side is the collision term due to
impurity scatterings, etc.
It has to be cautioned that the Boltzmann equation is valid only when
the scattering broadening of a MBB is small compared with its
bandwidth. This does not pose
essential difficulty when $p/q$ is a simple fraction.
In the more general situation of a large or infinite $q$ (irrational $\phi$),
we have to modify our approach in the following way: We divide the total
magnetic field $B$ into $B_0$ and
$\delta B$, where $B_0$ relates to the band structure not destroyed by
disorder,
and $\delta B$ is a small perturbation.  Then the semiclassical dynamics in the
MBBs of $B_0$, driven by $E$ and $\delta B$ (see Eqs.~(7) and (9) in the next
section),
will be employed in the Boltzmann equation.
Under such a circumstance, the scattering broadening is only required to be
smaller
than the bandwidth for $B_0$.

To demonstrate the use of Eq.~(3) in transport problems involving an electric
perturbation,
we consider a homogeneous system
in which $f$ depends on ${\bf k}$ only, and
use the relaxation time approximation.
The current to first order in ${\bf E}$ is:
\begin{eqnarray}
{\bf J}_n &=& {\bf E} \times  {\hat z}
\frac{e^2}{\hbar}\int \frac{d^2{\bf k}}{(2\pi)^2}f_0\Omega_n({\bf k})\\
\nonumber
&+& \left(\frac{e}{\hbar}\right)
^2 \int \frac{d^2{\bf k}}{(2\pi)^2}\tau({\bf k})
\left(-\frac{\partial f_0}{\partial {\cal E}}\right)\left({\bf E}\cdot
\frac{\partial {\cal E}_n}{\partial {\bf k}}\right )\frac{\partial {\cal
E}_n}{\partial {\bf k}},
\end{eqnarray}
where $\tau({\bf k})$ is the relaxation time and $f_0$ is the unperturbed
Fermi-Dirac distribution.
The first term is new and is due to the Berry phase curvature, which is nonzero
in general.
In fact, in the
simple case of a filled band ($f_0=1$), for which the second term is zero,
this term reduces to the topological Chern number
discovered by Thouless {\it et al} \cite{TKNdN}.
The second term is the usual Boltzmann transport formula \cite{Mermin}.
It was used for the calculation of longitudinal conductivity
of magnetic bands \cite{MacKinnon}.   Our theory justifies this usage
because the  Berry phase term only contributes to the Hall conductivity.

Compared to the brute force, all purpose Kubo formula approach, the
semiclassical dynamics, in conjunction with the Boltzmann transport theory,
offers a simple and intuitive picture of the behavior of the physical system.
The semiclassical dynamics also provides a useful tool in problems with
spatially varying and/or time dependent
fields, where quantum mechanical calculations are usually
very involved.  A detailed study will appear in a separate publication
\cite{long}.

\vskip .1in
\noindent{\it Magnetic perturbation and hyperorbits ---}
The remaining part of this letter will
focus on how the MBBs for a given magnetic field $B_0$ are perturbed by adding
$\delta B$.
In this case, the equations for the semiclassical dynamics become
\begin{equation}
{\bf{\dot r}}=\frac{\partial {\cal E}_n({\bf k})}{\hbar\partial {\bf k}}-
{\bf{\dot k}}\times {\hat z}\Omega_n({\bf k}),\ \
\hbar{{\bf\dot k}}=-e{{\bf\dot r}}\times \delta B{\hat z}.
\end{equation}
This result can be derived, for instance,
by considering a wave packet in a MBB and studying how its center of mass
moves in ${\bf r}$-space and ${\bf k}$-space \cite{long}.
Notice that the wave vector ${\bf k}$, which is a good quantum number
for $B_0$, is no longer conserved
in the presence of $\delta B$, even when both $B_0$ and $\delta B$ are uniform.

After ${\bf\dot r}$ is eliminated by combining both equations in Eq.~(7),
the equation for ${\bf\dot k}$ takes the following form:
\begin{equation}
\hbar{{\bf\dot k}}=
-\frac{\partial {\cal E}_n/\partial{\bf k}
\times{\hat z}\delta B e/\hbar}{1+\Omega_n({\bf k})\delta B e/\hbar}.
\end{equation}
It is not difficult to see that ${\bf k}$ moves along a constant energy contour
in the magnetic band structure.
The presence of $\Omega_n({\bf k})$ changes the speed of
motion, but it does not alter the shape of the orbit for a given energy.
The cyclotron orbit in ${\bf r}$-space can be derived from
${{\bf\dot r}}={{\bf\dot k}}\times {\hat z}(\hbar/e \delta B)$,
which shows that the ${\bf r}$-orbit is simply the ${\bf k}$-orbit
rotated by $\pi/2$ and scaled by the factor $\hbar/e\delta B$.
Such ``hyperorbits" were introduced by Pippard \cite{Pippard}.
We emphasize that, the existence of hyperorbits is a quantum effect, and cannot
be explained classically.
One possible way to detect them is by using an electron focusing device with a
configuration similar to a mass spectrometer \cite{Houten}.
In order to have a successful observation, the hyperorbit has to
be within the ballistic range of electron transport.

Analogous to ordinary cyclotron orbits, these hyperorbits will drift
in an external electric field.
By adding a $-e{\bf E}$ term to the second equation in Eq.~(7), we obtain
\cite{Mermin}
\begin{equation}
{\bf{\dot r}}=\frac{\hbar}{e\delta B}{\bf{\dot k}}\times {\hat z}-
\frac{{\bf E}\times {\hat z}}{\delta B}.
\end{equation}
The time average of the first term for a {\it closed} orbit is zero,
while the second term describes the drifting of the hyperorbit that results
in a Hall current. This will be used later to calculate the
Hall conductivity for magnetic subbands in the Hofstadter spectrum.

\vskip .1in
\noindent{\it Quantization of hyperorbits ---}
The dynamical equation for ${\bf k}$ in Eq.~(8) can be cast into the
Lagrangian formulation
\begin{equation}
L({\bf k},{\bf{\dot k}})=\frac{\hbar^2}{e\delta B}
(k_1{\dot k}_2-k_2{\dot k}_1)-{\cal E}_n({\bf k})
+\hbar{\bf A}_n\cdot{{\bf\dot k}},
\end{equation}
where ${\bf A}_n$ is the ``vector potential" for the Berry phase that
satisfies $\nabla\times {\bf A}_n({\bf k})=\Omega_n({\bf k}){\hat z}$.
Apart from an unimportant constant, the propagator for a completed closed
orbit $C$ in period $T$ is given by $\exp(i/\hbar \int_0^T L dt)$.
For a semiclassical orbit, the amplitudes for paths
that circle different times must add constructively.
This leads to the following quantization rule for the area of
a hyperorbit in
the $n$-th MBB \cite{Maslov},
\begin{equation}
\frac{1}{2}\oint_{C_m} ({\bf k}\times d{\bf k})\cdot{\hat
z}=2\pi\left(m+\frac{1}{2}
-\frac{\Gamma_n(C_m)}{2\pi}\right)\frac{e\delta B}{\hbar},
\end{equation}
where $m$ is a non-negative integer,
and $\Gamma_n(C_m)$ is the Berry phase for orbit $C_m$ \cite{quantization}.

By using the constraint that the area of
the outer-most orbit be smaller than the area of the first magnetic Brillouin
zone of size
$(2\pi/a)^2/q$,
we found the number of quantized orbits to be the integer part of
$1/(q\delta \phi)+\Gamma_n(C_{max})/2\pi+1/2$, where $\delta \phi=\delta B
a^2e/h$.
$\Gamma_n(C_{max})/2\pi$ can be replaced by the
Hall conductivity $\sigma_n$ (in units of $e^2/h$),
because the Berry
phase for the orbit $C_{max}$ is very close to $2 \pi\sigma_n$.
For an integer value of $1/(q\delta \phi)$, we then have
\begin{equation}
{\rm number\ of\ orbits\ }=\left|\frac{1}{q\delta \phi}+\sigma_n\right|.
\end{equation}
These orbits will be broadened into subbands by tunneling to orbits
with the same energy in other magnetic Brillouin zones.
Therefore, this naive-looking formula relates the Hall conductivity $\sigma_n$
of a
parent band to the number of daughter subbands
under a perturbation $\delta \phi$. It is crucial in understanding
the clustering pattern for the Hofstadter spectrum.

\vskip .1in
\noindent{\it The Hofstadter spectrum ---}
Consider a Bloch band subject to a magnetic flux
that can be expanded as
\begin{equation}
\phi={1\over\displaystyle f_1+
{\struct 1\over\displaystyle f_2+{\struct 1\over\displaystyle
{\struct f_3+\cdots}}}}.
\end{equation}
Its $r$-th order approximation will be written as $\phi_r=p_r/q_r$.
At the first order, the
Bloch band is broken into $f_1$ subbands; each subband is
further fragmented by an extra magnetic field at the second order, and so on.
As will be shown below, the new semiclassical dynamics offers a clear and
intuitive
picture about how each subband (parent) in the $r$-th order should
split into daughter subbands \cite{Niu}.

Our major findings are summarized below: (a)
Firstly, it is important to distinguish between ``closed" and ``open" subbands.
We define a ``closed" subband to be a subband broadened from a closed
hyperorbit;
similarly an ``open" subband is derived from an open orbit.
For a square or a triangular lattice,  all subbands except one
for every parent band are closed.
(b) The closed subbands at the same order all have the same Hall conductivity
\begin{equation}
\sigma_r = (-1)^{r-1}q_{r-1},
\end{equation}
where the subscript in $\sigma$ refers to the order, not to the band index
of the subband. If a parent band has only one open daughter band (eg.,
for a square or a triangular lattice), the Hall conductivity for
this open subband is $(-1)^{r-1}q_{r-1}+(-1)^rq_r$.
(c) Because of the difference in the Hall conductivities, a closed band will
break
into $f_{r+1}$ subbands at the next order, while an open band will break into
$f_{r+1}+1$ subbands.

We briefly describe the derivation for the Hall conductivity of a closed
subband.
It is determined by the response of the corresponding hyperorbit under
an electric field. Using Eq.~(9), we know that the drifting velocity of a
closed orbit is
$\langle {\bf\dot r}\rangle=-{\bf E}\times{\hat z}/{\delta B}$. It follows that
the Hall conductivity for a closed subband at the $r$-th order
is $\sigma_{r}=e\rho_r/\delta B_{r-1}$, where
$\rho_r$ is the electron density per unit area and ${\delta B}_{r-1}=h\delta
\phi_{r-1}/(ea^2)$.
Since the electrons are equally distributed among the $q_{r}$ subbands at the
same order,
$\rho_r$ is equal to $1/q_{r}$ times
the electron density of the original Bloch band.
Eq.~(14) is obtained after the identity $p_{r}/q_{r}-p_{r-1}/q_{r-1}=
(-1)^{r-1}/(q_{r}q_{r-1})$ is used to evaluate $\delta\phi_{r-1}$.
The Hall conductivity for an open subband can be figured out quite easily
by using the sum rule: $\sigma_{\rm parent}=\sum \sigma_{\rm daughter}$
\cite{long}\cite{Avron}.

For a square lattice, an open daughter band is
always located at the center of a parent band; therefore, we know
which subband the conductivity $\sigma^{\rm open}$ belongs to.
The Hall conductivity distribution obtained this way is exactly
the same as that which is obtained by using the Diophantine equation with some
subsidiary constraints
\cite{TKNdN}. We should emphasize that
this is the first time the (seemingly) erratic behavior of the Hall
conductivities for the
Hofstadter spectrum is given a clear and direct physical meaning.

With the help of Eq.~(12), we can determine how a parent band is splitted by
$\delta \phi$.
Substituting $\delta \phi_{r}=(-1)^r/(q_{r+1}q_r)$ into Eq.~(12),
and taking into account the $q_r$-fold degeneracy for the $r$-th order
magnetic Brillouin zone,
we then have the number of daughter bands for an $r$-th order parent band
\begin{equation}
{\cal D}_{r}=|(-1)^rq_{r+1}+\sigma_r|/q_r.
\end{equation}
In conjunction with Eq.~(14), we obtain ${\cal D}_{r}=f_{r+1}$ for a closed
band.
Similarly we have ${\cal D}_{r+1}=f_{r+1}+1$ for an open band.
Azbel conjectured that there are $f_{r+1}$ daughter subbands for {\it every}
parent band \cite{Azbel}. It is clear from our calculation
that his conjecture is correct only for a closed parent band.
An expression similar to Eq.~(15) has been obtained by Wilkinson, but his
evaluation
of such an expression required external input for the Hall conductivity
\cite{RG}.

Various values of $\phi$ have been used to check the predicted splitting of
subbands
with the actual Hofstadter spectrum,
and the results are found to agree very well.
One exception occurs when there is a degeneracy among the subbands.
In this case the individual $\sigma_r$, as well as ${\cal D}_{r}$, cannot be
determined uniquely.
For example, when $\phi=1/4$, the central two subbands are degenerate for a
square lattice
and the Hall conductivities can be either $(1,1,-3,1)$
or $(1,-3,1,1)$ ($q_0=1$).
One of these two subbands will have this $-3$ ($\sigma^{\rm open}$)
when the degeneracy is lifted by
next nearest neighbor coupling \cite{Hatsugai}. As expected, we found the
subband
corresponding to a hyperorbit that is closer to the nesting energy contour
acquires
this $\sigma^{\rm open}$.

For a triangular lattice, the open orbit is near the zone boundary
(exact location requires some calculation). Therefore,
the distribution of Hall conductivities and band splitting
are no longer symmetric in energy.
This  explains the asymmetry in the spectrum generated
by numerical calculation \cite{Hatsugai}.
For a lattice without 3-fold or 4-fold symmetry, more than one open orbit
may exist in a range of energy. In this case, the total Hall conductivities
for open orbits can be figured out by the sum rule. However,
further analysis is required to get the detailed distribution within them.

In summary, we have demonstrated the new semiclassical dynamics for  magnetic
Bloch bands
and its application to a variety of phenomena involving strong magnetic fields.
It can be used to calculate the transport properties,
to obtain the quantization rule for hyperorbits, and
to get the Hall conductivity for a magnetic subband.
We also showed that the complex structure of the Hofstadter spectrum can be
explained
as a logically consistent part of this theory.

\acknowledgments
Q.~N. wishes to thank F.~H.~Claro, M.~Kohmoto, W.~Kohn, and M.~Marder
for many helpful discussions.
M.~C.~C. wishes to thank E.~Demircan, G.~Georgakis, and R.~Janhke for their
comments
on the manuscript. This work is supported by the R.~A.~Welch foundation.

\end{document}